\documentclass[fleqn,12pt,twoside]{article}
\usepackage{espcrc1}

\usepackage[final]{graphicx} 

\title{Short-ranged attractive colloids: What is the gel state ?} 
\date{\today} \author{E.~Zaccarelli\address[roma]{Dipartimento di 
Fisica and INFM Center for Statistical Mechanics and Complexity, 
Universit\`a di Roma ``La Sapienza'', Piazzale Aldo Moro 2, 00185 
Roma, Italy}, F.~Sciortino\addressmark[roma], 
S.V.~Buldyrev\address{Center for Polymer Studies and Department of 
Physics, Boston University, Boston, MA 02215, USA.}, and 
P.~Tartaglia\addressmark[roma]} 
\begin{document} 
\maketitle 
\begin{abstract} 
We evaluate thermodynamic, geometric and dynamic properties of a 
short-ranged square well binary mixture to provide a coherent picture 
of this simple, but rich, model for colloidal interactions.  In 
particular, we compare the location, in the temperature-packing 
fraction plane, of the geometrical percolation locus, the metastable 
liquid-gas spinodal and the glass transition lines.  Such comparison 
provides evidence that the gel-state can not be related to the 
attractive glass transition line directly. Indications are given for 
the possibility of an indirect link between the two, via an arrested 
phase separation process.  We finally discuss the possibility that a 
spherical short range attraction may not be sufficient to produce an 
equilibrium cluster phase at low packing fraction and low 
temperatures. 
\end{abstract} 
%\pacs{64.70.Pf, 82.70.Dd} 

\section{Introduction}

Colloidal dispersions are a suitable class of matter for many 
scientific purposes. Indeed, these systems are experimentally accessible with 
light scattering techniques and microscopy, due the the large length 
scales and time scales involved. Also, the inter-particle interactions 
can be tuned almost {\it ad-hoc}, for example by changing the solvent, 
grafting the particles or adding polymers in the dispersion. 
Interaction ranges much shorter than the characteristic ones of 
molecular or atomic liquids can be produced in colloidal 
suspensions. The  interaction range can be reduced  to a 
few per cent of the colloidal particle diameter. Hence, colloids can be 
used to test a large variety of theoretical models, or vice versa 
theories can be subjected to stringent experimental tests. 
Experimentally, the most common realization of short-range attractive 
colloids is obtained by the addition of small non-adsorbing polymers 
in the colloidal solution. These, for sufficiently small sizes, can be 
integrated out of the description\cite{likos}, and their net effect 
is to produce an effective attraction on the colloidal spheres, via 
depletion interactions\cite{AO}. The size and the concentration 
of the polymers control respectively the range and the magnitude of 
the attraction. Theoretical models of an effective one-component 
attractive potential are often used, to mimic the experimental 
situation.

The thermodynamics of short-ranged attractive colloidal systems has
been studied in great details\cite{frenkel,lekker}.  By tuning the
inter-particle interactions, it has been shown that, for spherical
hard-sphere colloids, the addition of a particularly narrow range of
the attractive part of the potential, with respect to the hard-core
diameter, can produce an interesting modification of typical ({\it \`a
la} Van der Walls) equilibrium phase diagrams.  On decreasing the
range of interaction of the attractive potential, the liquid-gas
coexistence curve becomes metastable with respect to the crystal,
resulting in the disappearance of the liquid as equilibrium phase.

In recent years, the study of the dynamic behaviour of short-ranged
attractive colloids has revealed exciting new results.  Thanks to
theoretical work based on the Mode Coupling Theory (MCT) for the glass
transition \cite{Goetze1991b}, applied to Baxter sticky spheres
\cite{Fabbian1999} and to the short ranged attractive Yukawa
model\cite{Bergenholtz1999}, some previous astonishing experimental
results\cite{oldpoon,oldbartsch} have been revisited and
interpreted. MCT theoretical predictions have provided a coherent
picture \cite{Dawson2001,mechanical,foffi,merida,messina,sperlpisa} of
the dynamic features characteristic of short-ranged attractive
potentials and have stimulated novel experiments
\cite{mallamace00,pham02,eckert02,malla02,malla03,pham2003pre} and
simulation studies
\cite{puertas02,foffi02,zaccarelli02,Zaccarelli2003,puertas03,Zaccarelliprl,Sciortino2003}.
Near the structural arrest dynamics displays far more richness than
the one observed in simply repulsive, or long-ranged attractive
systems. Many predictions have been confirmed by multiple evidences,
some have been questioned, others are still under investigation.

So far, most of the attention of the scientific community has
addressed questions regarding the behaviour of short-ranged attractive
colloidal systems in the very dense regime of colloidal particles,
where the most striking predictions of MCT are manifested. In
particular, it is now recognized, with the help of experiments and
numerical simulations, that, at high volume fractions, two different
mechanisms for glassification exist, one controlled by the excluded
volume, commonly referred as `hard-sphere' (repulsive) glass
transition, and one dominated by the attractive interactions, or
`bonding', between particles, commonly termed `attractive' glass
transition. The former glass scenario is only observed at high packing
fraction $\phi$, while the latter is manifested for large strength of
attraction, i.e. low temperatures or, in the depletion picture, large
polymer concentration.  These two mechanisms effectively compete with
each other\cite{natmat}, when the range of the attractive part of the
potential becomes sufficiently short with respect to the hard-core of
the particles, giving rise to a reentrant fluid region between the two
glasses. The system remains liquid even for packing fractions where a
hard sphere repulsive system would be glassy.  In MCT, this
competition arises from the presence of a higher-order singularity in
the control-parameter space \cite{sperl02,sperl03}, which regulates
the anomalous dynamical behaviour in these systems, giving rise to an
intriguing logarithmic decay of the density auto-correlation
functions, as well as a power-law sub-diffusive behavior for the mean
squared displacement. For the square well model, a recent study has
provided evidence of the presence of such point\cite{Sciortino2003}.
 
However, not all predictions share the same robustness with respect to 
the approximations present in the MCT description. In particular, the 
theory is {\it ideal}, in the sense that it does not take into account 
the so-called hopping processes, that, as well known, become relevant 
on approaching the glass transition. These processes allow for 
residual diffusion where the theory would predict a complete 
arrest. Indeed, a recent work \cite{Zaccarelliprl} has 
investigated the question of the existence of a pure glass-glass 
transition, as well as the stability of the attractive glass, in 
relation to activated bond-breaking processes. 
 
Experimentally, dynamical arrest phenomena in short range attractive
colloids are observed not only at high density, as discussed above,
but also in the low packing fraction region.  In this case, the
arrested material is commonly named a gel.  In some cases, the gel
phase appears to be contiguous to a cluster
phase\cite{Segre,FaradayWeitz,Dinsmore,lysozime}, characterized by
large supraparticular aggregates, diffusing through the sample in
ergodic dynamics.  The gel state displays peculiar features like the
appearance of a peak in the static structure factor, for very large
length scales (of the order of several particle diameters), that is
stable in time, as well as a non-ergodic behaviour in the density
correlation functions and a finite shear modulus.  These solid-like,
disordered, arrested features have induced to the appealing conjecture
that these colloidal gels can be viewed as the low-density expression
of the attractive glasses, both being driven by the same underlying
mechanism of arrest.  Support to these ideas was found in MCT
itself. Indeed, the theory predicts that the attractive glass line
extends, practically flat, toward very low packing fractions, almost
touching at the critical point the spinodal curve, and following it on
its left-hand side.  The predicted large values of the non ergodicity
parameter along the attractive glass line, similar to the ones
observed in the gel phase, and the possibility of modeling the ergodic
to non-ergodic transition locus with the attractive MCT line provide
support for such an identification
\cite{Bergenholtz1999,langmuir}. Still, some inconsistencies in the
interpretation of the gel state as attractive glass have been
noted. For example, in the low packing fraction regime, an
identification of the attractive glass line with the gel line theory
would predict a gel made of particles with average number of bonds
less than two \cite{mechanical}.
 
In this paper, we present Molecular Dynamics simulations of a simple 
model of short-ranged potential.  In the high packing fraction region, 
the dynamics of the model displays two different arrest mechanisms, 
and indeed two different glass transition lines have been 
located. Here we complement the high density dynamic data with an 
evaluation of the percolation locus and of the liquid-gas spinodal. 
We study the intersection between the glass line and the spinodal and 
show that the former meets the latter at low 
temperatures, on the high density side. Similarly, no relation is 
found between the percolation locus and the glass 
line\cite{russel,douglas}. The outcome of our studies is quite 
unexpected, and might be crucial for understanding the nature of the 
colloidal gels. 
 
\section{Model, Theory and Simulation} 
 
We perform Molecular Dynamics simulations of a $50\%-50\%$ binary
mixture of $N=700$ hard spheres of mass $m$, with diameters
$\sigma_{_{AA}}$ and $\sigma_{_{BB}}$ and ratio
$\sigma_{_{AA}}/\sigma_{_{BB}}=1.2$. The hard core between particles
of different type $\sigma_{_{AB}} = 0.5
(\sigma_{_{AA}}+\sigma_{_{BB}})$. The hard core potential is
complemented by an attractive square well potential of depth $-u_0$,
whose well-width $\Delta_{ij}$ is controlled by varying the parameter
$\epsilon=\Delta_{ij}/(\Delta_{ij}+\sigma_{ij})$, which is the same
for any $i,j$.  In this model, a `bond' between two particles is
unambiguously defined when their interaction energy is equal to
$-u_0$.  The simulation is based on a standard event-driven algorithm,
implemented for particles interacting with SW
potentials\cite{Rapaport97}. Distances are measured in units of
$\sigma_{BB}$, while energy and temperature are measured in units of
$u_o$, i.e. $k_B=1$. Time is measured in units of
$\sigma_{BB}\cdot(m/u_0)^{1/2}$, mass in units of $m$.
 
For the particular case $\epsilon=0.03$, results of extensive 
simulations for this model have been reported in the high density 
region, in order to make comparison with the existing MCT predictions. 
This was carried forward in a recent attempt to build an effective 
mapping between theory and simulation, in order to localize and probe 
the dynamics near the so-called higher order singularity\cite{Sciortino2003}. 
 
Ideal glass lines are predicted by MCT as ergodic to non-ergodic
transitions, at which by definition the self-diffusion coefficient of
atoms, or colloidal particles in our case, is equal to zero. In order
to build from our simulations something comparable to an ideal glass
line, we have calculated iso-diffusivity curves in the $\phi - T$
plane and studied the evolution of these curves with decreasing
diffusivity. Interestingly, the shape of the ideal glass line is
maintained up to quite large values of the diffusivity, allowing for a
straightforward establishment of the re-entrant behaviour of the glass
transition.
 
Combined with the iso-diffusivity curves, we present here the 
estimation of liquid-gas `pre-critical' curve. Such curve is defined 
as the locus in the $\phi - T$ plane of points where the static 
structure factor $S(q) \sim 1$ at $q=0$ and provides a close estimate 
to the spinodal line. Indeed, $S(0)=1$ signals the onset of the 
divergence of the compressibility, i.e. it is a precursor of the 
phase separation into gas (colloid-poor) and liquid (colloid-rich) 
phases. 
 
The square well model is also very well suited for defining a
percolation threshold. Indeed, as discussed above, the existence of a
bond can be defined unambiguously, when the pair interaction energy is
$-u_0$.  We report here the bond percolation line, beyond which
space-spanning clusters of bonded particles are present in the system
at a given instant. We estimated it by calculating the cluster
connectivity. From an operational point of view, we have defined
percolating a state point where more than 50\% of the examined
independent configurations (over a total of sixty) displayed an
infinite cluster, spanning across the simulation box.
 
\section{The Phase Diagram}

\begin{figure} 
\begin{center}  
\includegraphics[width=12cm,angle=0.,clip]{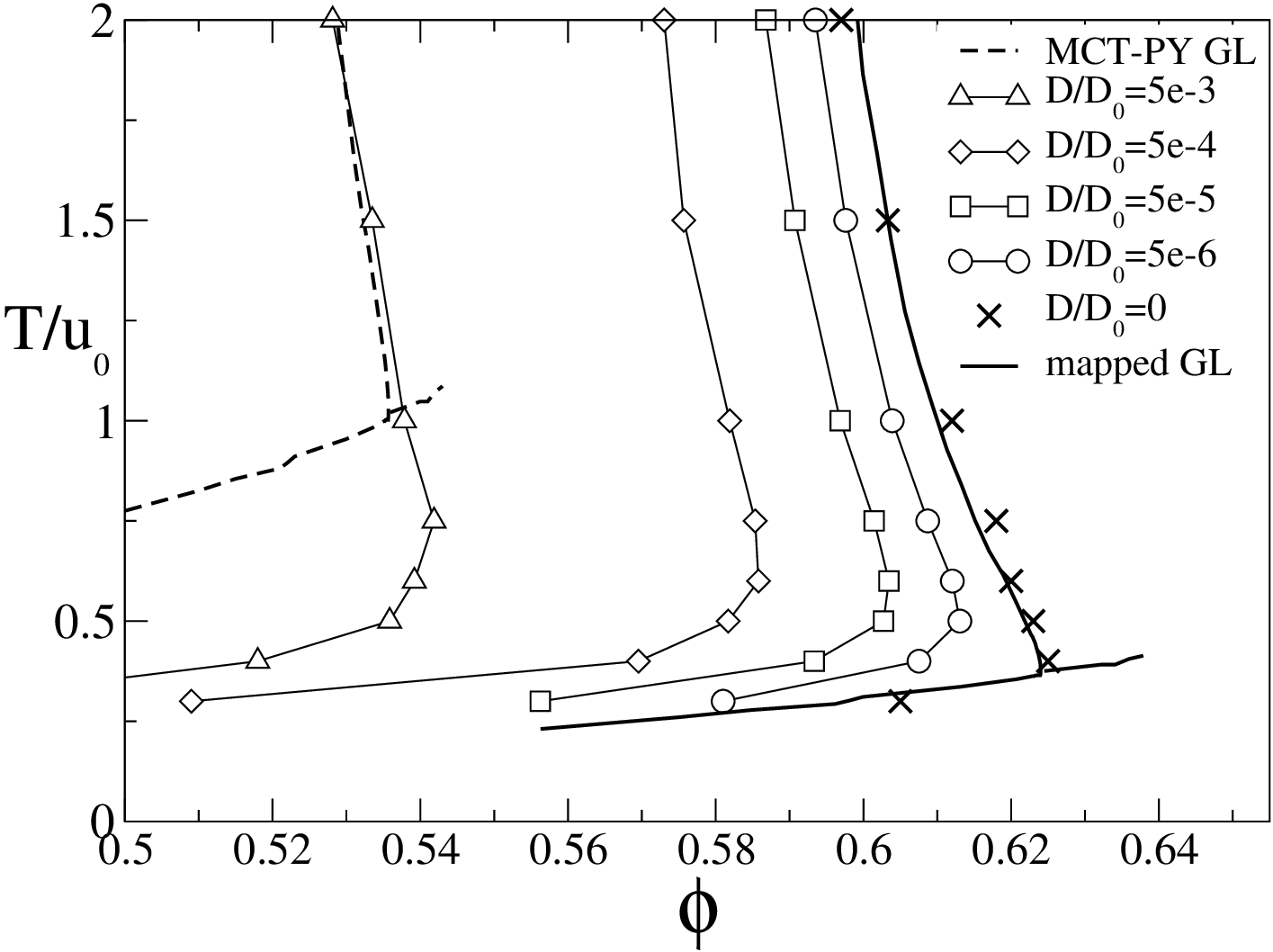} 
\caption{$(\phi, T/u_0$)-diagram in the large colloidal volume
fraction region for the binary SW system with $\epsilon=3\%$. From
left to right the reported curves are: the calculated MCT glass line
(GL) within Percus-Yevick approximation, iso-diffusivity curves with
$D/D_0$ equal to $5\cdot10^{-3}$, $5\cdot10^{-4}$, $5\cdot10^{-5}$ and
$5\cdot10^{-6}$, power-law extrapolated MD data from
Ref.\protect\cite{zaccarelli02} for $D/D_0=0$ (crosses), and the MCT
GL mapped as shown in Ref. \protect\cite{Sciortino2003}.}
\label{fig:fig1a} 
\end{center} 
\end{figure}

We plot in Fig.\ref{fig:fig1a} the ideal MCT glass lines (GL) for the 
studied binary SW system with $\epsilon=3\%$ calculated theoretically 
using the Percus-Yevick (PY) structure factor\cite{Sciortino2003}. We 
also report the locus of constant diffusivity, evaluated from MD 
simulation, with $D/D_0$ \cite{notaD0} values varying between $5\cdot 
10^{-3}$ and $5\cdot 10^{-6}$ together with the locus of zero 
diffusivity (symbols), obtained extrapolating isothermally the 
$\phi$-dependence of the diffusion coefficient according to 
power-law\cite{Sciortino2003}.  We also show the transformation of the 
ideal PY MCT glass line according to the bilinear 
transformation\cite{sperl03} $\phi \rightarrow 1.897\ \phi -0.3922$ 
and $T \rightarrow 0.5882\ T - 0.225$, already discussed in 
\cite{Sciortino2003}.  The parameters of the transformation are the 
result of fitting the locus of zero diffusivity with the transformed 
MCT curves.  Indeed, it is typical that both MCT and the PY solution, 
in a certain sense, `overestimates' the glass transition, i.e. always 
predicting it earlier with respect to reality (for example at 
higher temperatures in super-cooled liquids or at lower packing 
fraction for hard spheres). However, differently from hard sphere 
systems, where this discrepancy requires just a shift in the packing 
fraction, for an attractive system we have to consider a bilinear 
mapping \cite{nota}, which considerably affects the attractive branch 
of the predicted glass line. 
 
\begin{figure} 
\begin{center}  
\includegraphics[width=12cm,angle=0.,clip]{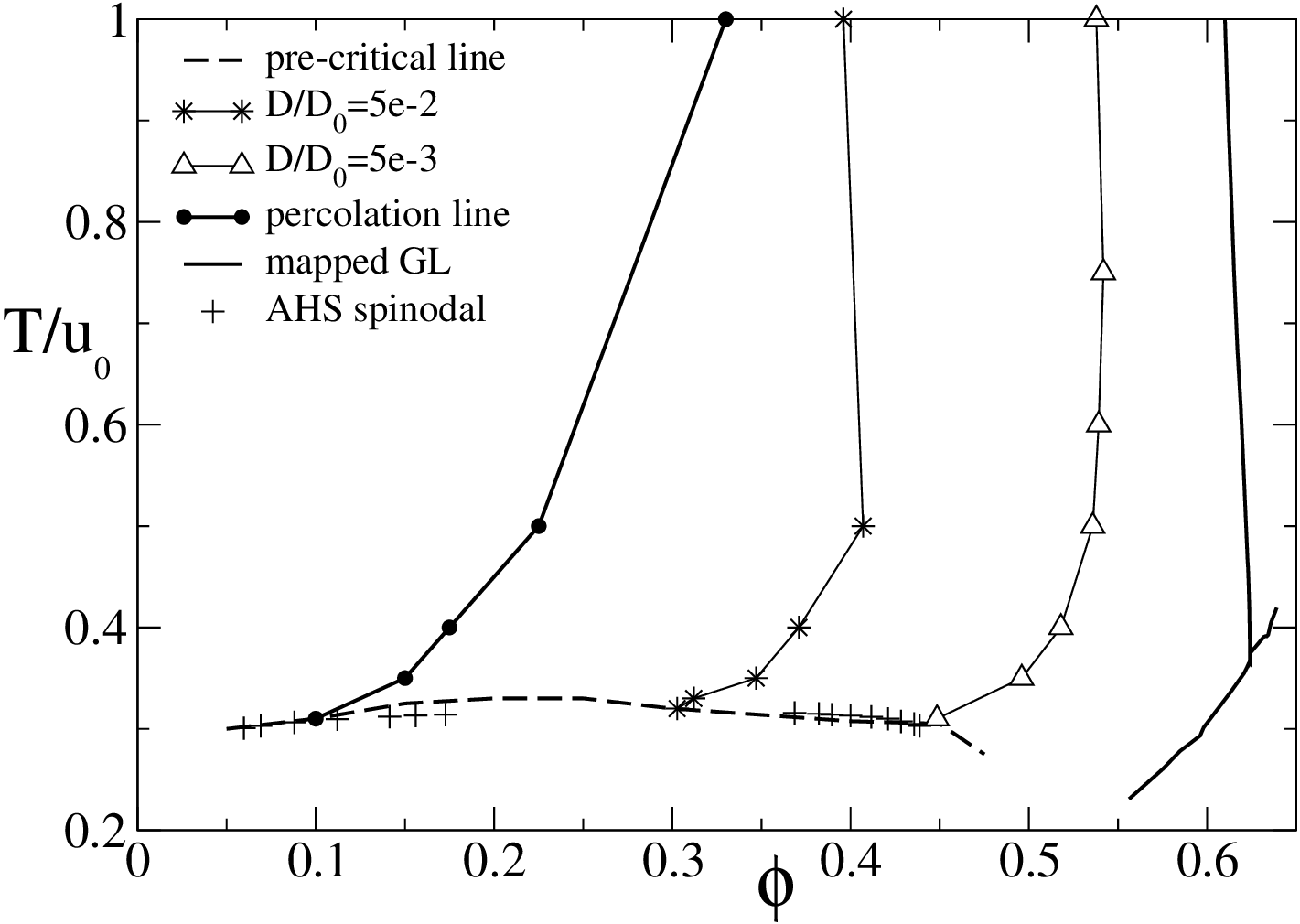} 
\caption{Same as Fig. \protect\ref{fig:fig1a}, but reporting the full
colloid density region. Together with the mapped GL, already shown in
the previous figure, the calculated percolation and precritical lines are
reported, as well iso-diffusivity curves for the values
$D/D_0=5\cdot10^{-2}, 5\cdot10^{-3}$, both crossing the spinodal line
at finite temperature. From these data, one could extrapolate the
temperature where the GL meets the spinodal $T_{cg}$ (see text). Also
shown for comparison are spinodal points calculated for the Adhesive
Hard Sphere (AHS) model, redrawn from \protect\cite{Miller2003}. }
\label{fig:fig1} 
\end{center} 
\end{figure}

Fig.\ref{fig:fig1} shows the `spinodal' curve, the percolation curve
and the locus of constant diffusivity with $D/D_0$ values of $5\cdot
10^{-2}$ and $5\cdot 10^{-3}$, significantly extending to lower
packing fractions the calculations already presented in
\cite{zaccarelli02}).  From the data shown in Fig.\ref{fig:fig1}
several considerations aris:\\ (i) the intersection between the
isodiffusivity curve and the spinodal provides an estimate of the
characteristic times along the spinodal.  Dynamics slows down on
increasing density.  The intersection between the glass line and the
spinodal is located on the high density side. Hence, in this model,
the attractive glass line cannot be {\it directly} associated to
physical phenomena taking place at low packing fractions, i.e. to the
gel state.  \\ (ii) The fact that the ideal attractive GL meets the
spinodal at high densities suggests an {\it indirect} possibility for
linking the gel state to the attractive glass.  Indeed, if we call
$T_{cg}$ the temperature at which the attractive GL meets the
coexistence line in the high density side\cite{nota2}, then quenches
below $T_{cg}$ may generate upon decomposition regions where the
particle concentration is within the attractive glass phase and hence
which could arrest kinetically the phase separation process, leaving
the imprinting of the phase separation in the frozen structure factor
of the
system\cite{sciortinoyoung,jackle,bibette,faradaypoon,lysozime}.  Such
hypothesis, discussed in more details in the next section, although
stimulating, unfortunately cannot explain the existence of a
contiguity between an "equilibrium" cluster phase and the colloidal
gel\cite{Segre,FaradayWeitz,Dinsmore,lysozime}.\\ (iii) The static
percolation line is found to start from the low density side of the
spinodal, and at all studied temperatures, it remains well to the left
side of the largest drawn iso-diffusivity curve. This means that the
percolating clusters are made of particles which are moving fast, thus
the lifetime of the bonds of which the clusters are made at the
percolation threshold is extremely short.  A study of the lifetime of
the infinite cluster\cite{youngsciortinoprl}, which may provide more
precise indication of the time stability of the percolating cluster as
compared to diffusional times is underway. Still, the short lifetime
of the bonds and the extremely large diffusional times suggest that it
is not possible to establish any connection between percolation and
formation of stable aggregates.  On the contrary, we can rule out the
possibility that, for the short ranged $\epsilon=3\%$ model studied
here, bond percolation is connected to gelation.\\
 
Recently, Miller and Frenkel \cite{Miller2003} evaluated the critical
point, the spinodal line and the percolation locus for the sticky
spheres Baxter model. The Baxter model is the limiting case of the
square well model when the width of the well goes to zero and $u_0$
goes to infinity. The relation between the sticky parameter $\tau$
\cite{baxter} and the square well parameters, i.e.  $u_0/k_B T=
log(1+1/(4 \tau (1/(1-\epsilon)^3-1)))$ , based on the equality of the
second virial coefficients for the two models \cite{noro}, and between
the Baxter packing fraction $\phi_B$ and the square well one, i.e.
$\phi=\phi_B(1-\epsilon)^3$, allow us to compare the phase diagram of
the two systems.  Miller and Frenkel \cite{Miller2003} data, upon
appropriate scaling of the variables, are also reported in
Fig.\ref{fig:fig1}. The agreement between the pre-critical curve and
Baxter spinodal allows us to estimate the critical point for the
attractive well to be approximately around $\phi_c \sim 0.24, T_c/u_0
\sim 0.31$ for $\epsilon=3\%$.


\section{Phase Separation and Gels} 
 
We now examine the possibility discussed in the previous section that
the gel state is an arrested phase separated system, where the high
density phase concentration is in the glass phase.  We perform such a
study by quenching from $T/u_0=1.0$ within the liquid-gas unstable
region.  We focus for simplicity on a one-component system but, since
now we are interested in an aggregation process which involves
clustering of many particles, we consider a much larger system than
the one studied previously.  We study $N=30000$ particles, interacting
via SW potential, with varying well-width, respectively equal to
$\epsilon=0.04,0.01,0.005$.
%For all these parameters we expect the locations of spinodal and percolation curves to be just weakly shifted, with respect to the ones reported in Fig.\ref{fig:fig1}, because for the limiting case of adhesive hard spheres studied in \cite{Miller2003}, results are very close to the ones of our binary mixture, and most importantly, they display the same overall features. 
 
We focus on the low packing fraction $\phi=0.15$, and decrease the
temperature. The choice of parameters was made in order to compare our
simulation data with the phase diagram reported in
\cite{FaradayWeitz}. Up to approximately $T/u_0=0.3$, the system
remains in equilibrium, and the static structure factor is similar to
that of a normal liquid. However, as we go further in lowering the
temperature, i.e. quenching the system to $T/u_0=0.2$ and $T/u_0=0.1$,
we enter in the spinodal regime and phase separation takes place. This
causes particles to aggregate, producing a low wavevector peak in the
static structure factor, similar, in localization and amplitude, to
the one observed in \cite{Segre}. However, differently from what
reported in those experiments, the structure factor continues to
evolve, during the simulation, although on logarithmic
time-scales. This phenomenon is more marked for the largest studied
well-width, i.e. $\epsilon=0.04$, and for the highest temperature
considered for the smaller widhts, $T/u_0=0.2$.
 
\begin{figure} 
\begin{center}  
\includegraphics[width=12cm,angle=0.,clip]{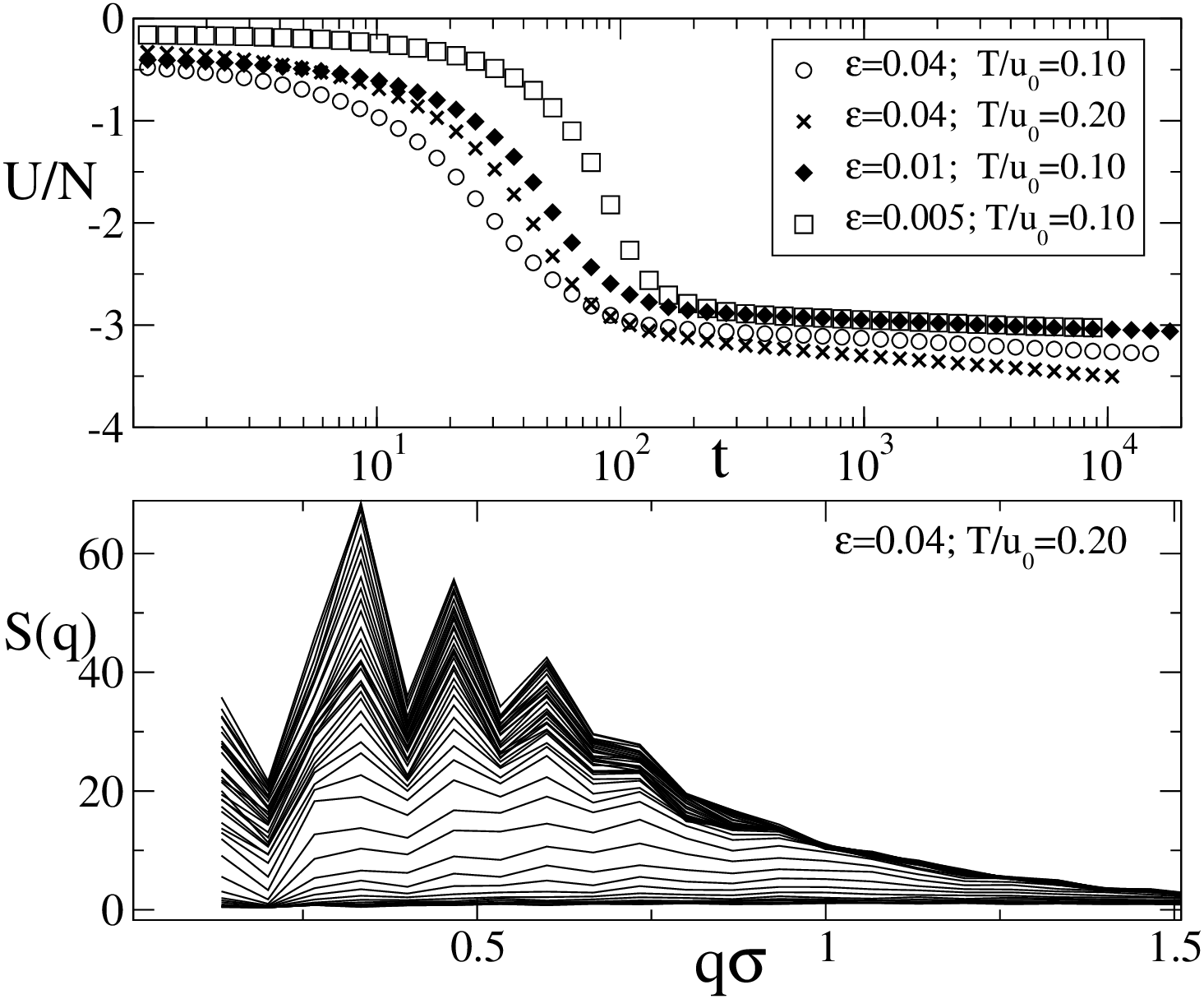} 
\caption{Upper Panel: Time evolution of the energy per particle for a
one-component SW model at $\phi=0.15$,after the quenches to
$T/u_0=0.2$ and $T/u_0=0.1$. Various well-widths; Lower Panel: Static
structure factors after the quench for the case $\epsilon=0.04$ and
$T/u_0=0.2$. The various curves refer to different times, equally
spaced on a logarithmic scale. The significant noise is due to the
fact that we show a single realization.}
\label{fig:fig2} 
\end{center} 
\end{figure}

\begin{figure} 
\begin{center}  
\includegraphics[width=12cm,angle=0.,clip]{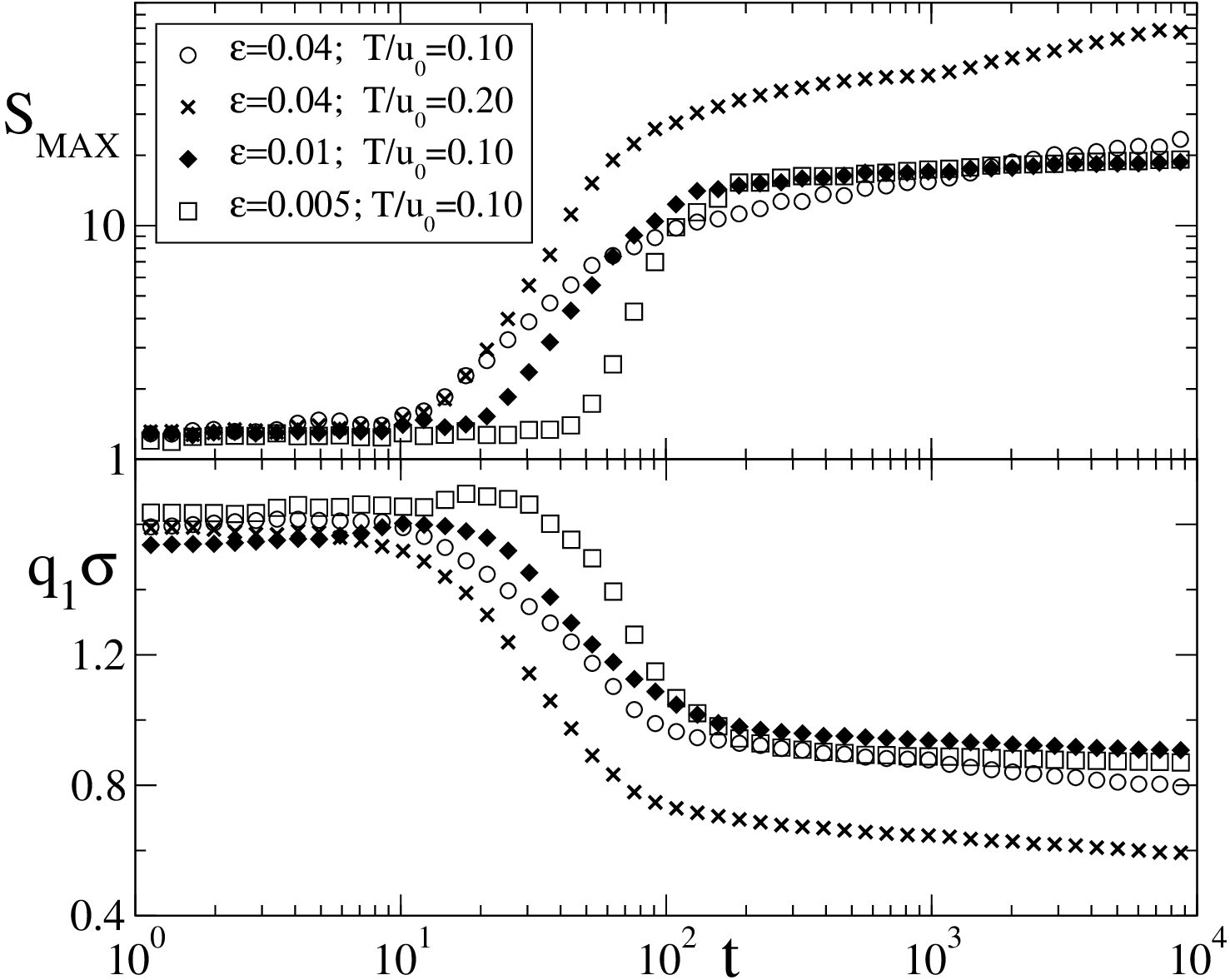} 
\caption{Upper Panel: Time evolution after the quenches of the static
structure factor maximum $S_{MAX}$, for the same cases as in
Fig. \protect\ref{fig:fig2} (upper panel); Lower Panel: Time evolution
after the quenches of the first moment $q_1$ in units of $\sigma$ of
the static structure factor. }
\label{fig:fig3} 
\end{center} 
\end{figure}

In Figure \ref{fig:fig2} - (upper panel), we report the time evolution
of the energy per particle $U/N$ for the various studied cases. It can
be observed that there is a characteristic time-scale, which controls
the aggregation kinetics. After the microscopic time-scale, there is a
strong decrease in the energy, which then crosses over to a regime of
very slow, approximately logarithmic, decay. The decay is slower, the
narrower the studied well-width and the lower the temperature. We note
that the number of bonds per particle tends to a value of about $3$ or
more, which, taking into account the low packing fraction of the
system, indicates the formation of a cluster network.  The resulting
cluster phase is not an equilibrium phase, in the sense that it is
driven by spinodal decomposition. To support such interpretation we
show in Figure\ref{fig:fig2} - (lower panel), the evolution of the
static structure factor for the case $\epsilon=0.01$ and
$T/u_0=0.1$. The various curves represents $S(q)$ at different times
from the quench, which are logarithmically spaced between $t_1\sim 10$
and $t_2 \sim 10^4$. Times shorter than thermal equlibration are not
reported. To better quantify the time dependence of the separation
process, we report in Figure \ref{fig:fig3} the time evolution of the
peak intensity $S_{MAX}$ (upper panel), in analogy with the inset of
Fig.~1 in Ref.\cite{Segre}. For the larger well-width $\epsilon=0.04$,
the increase of the amplitude in $S(q)$, although on logarithmic
time-scale, is still clearly observable after two decades in
time\cite{heavy}. However, for the two very narrow widths $1\%,0.5\%$
at the extremely low temperature $T/u_0=0.10$ a significantly flatter
behaviour is observed. Indeed, the two curves are almost superimposed
onto each other at long times, and the narrower well-width shows a
sharper crossover to the {\it quasi-plateau}.


Also, in Figure \ref{fig:fig3} - (lower panel), we plot the time 
evolution of the first moment of $S(q)$, i.e. $q_1=[\sum_q q 
S(q)]/[\sum_q S(q)]$, which has a scaling equivalent to the peak 
position, but which can be calculated much more accurately 
\cite{glotzer94}.  These curves can be fitted at large times with a 
power-law $\sim t^{-\beta}$, where $\beta$ varies between $0.25$ and 
$0.5$, depending on the case and on the range of fit.  We recall that 
the typical exponent for spinodal decomposition is $1/d$, $d$ being 
the dimensionality \cite{spinodal,bibette}. This 
result is consistent with the typical scaling of spinodal 
decomposition, and provides another indication that the aggregation we 
observe is driven by the gas-liquid phase transition.  The two cases of 
extremely short-ranged and strong attraction display an extremely slow 
dynamics, which almost arrest at large times. If we return to the 
situation examined in the previous paragraph, we know that the 
attractive branch of the glass line meets the spinodal, at some very 
low temperature $T_{cg}$. It might be possible that these two cases 
already correspond to a temperature lower than $T_{cg}$, thus what we 
observe here are {\it indirect} gels induced by spinodal 
decomposition.

\section{Conclusions} 
 
In this paper we have provided evidence of two new important facts for 
simply short-ranged attractive potentials, of the kind generated via depletion 
interactions.  Firstly, the slowing down of the dynamics in 
short-ranged attractive systems truly arises only at very high 
densities, far away from the percolation line.  At low temperatures, 
the attractive glass line ends in the spinodal {\it on the high 
density side}.  Secondly, a gel contiguous to an equilibrium cluster 
phase does not manifest. A gel could result from an 
arrested phase separation process when the density of one of the two 
phases crosses the attractive glass transition at low temperatures and 
for very short ranges of attraction ($T/u_0=0.1$, $\epsilon \le 
0.01$).  At even lower temperatures, diffusion limited cluster 
aggregation\cite{dlca} will take over, leaving the 
imprinting of the aggregation process in the gel 
structure\cite{carpineti,stold}. 
 
These two observations clash with earlier conjectures that a gel
state, considered as a natural extension to lower densities of the
ideal MCT attractive glass, would be found in these systems
\cite{Bergenholtz1999,langmuir}. We also note that the data reported
in this article appear in disagreement with the recent theoretical
model \cite{clusterMCT} (see in particular Fig. 2 of
Ref. \cite{clusterMCT}), where a gel state is identified with an
attractive glass after a renormalization procedure.  On the other
hand, our results shed some light on the fact that in real systems
there must be an additional mechanism that allows to stabilize an
equilibrium cluster phase.  Evidence is now emerging in the scientific
community that colloidal particles often tend to have a residual
charge distribution \cite{bladereen,Dinsmore,poon}, whose net effect
is a weak, long-ranged repulsive barrier, whose importance in
governing the dynamics of the system may manifest when many particles
are clustered together, preventing further aggregation\cite{kegel}.
%We want to remark that this weak repulsion can also be tuned experimentally. However, standard experiments with sterically stabilized colloidal spheres in solution with non-adsorbing polymers do contain a small amount of charge, which should not consistently affect the equilibrium phase diagram of the simply attractive models. On the other hand, in the potential discussed in\cite{Puertas2002} a large barrier is introduced to stabilize the system against phase separation, thereby changing considerably the equilibrium picture of that model with respect to the one described here. 
It would be interesting to find out if the stabilizing effect of 
electro-static forces implies that gels are formed by caging of 
clusters, instead of by caging of particles. If this were the case, 
particle gels would be driven by a completely different mechanism than 
attractive glasses, ultimately being a different form of  
glasses, where particles are replaced by clusters of many 
particles, whose size and distribution will depend on the amount of 
charge that is present in the colloidal suspension.  Work is in progress to 
address specifically such possibilities.

\section{Acknowledgments} 
 
We acknowledge support from MIUR PRIN, FIRB and 
INFM PRA-GENFDT, and also from INFM Iniziative Calcolo Parallelo. 
We thank W. C. K. Poon for interesting discussions.


\begin{thebibliography}{99}
\bibitem{notaD0} We recall that $D_0= \sigma \sqrt{T/m}$, is a 
normalizing factor, which ensures that the difference in the average 
velocities due to the temperature is eliminated, so that the diffusion 
coefficient is comparable between different temperatures.
\bibitem{nota} This mapping is based on the hypothesis 
that the parameter $\epsilon$ does not require any transformation, 
based on the observation that it is the crucial parameter governing 
the anomalous behaviour of the system and the existence of the higher 
order singularity.
\bibitem{nota2}  The ideal GL, being the 
outcome of the ideal MCT does not coincide with the "calorimetric" 
glass transition line. While in hard-spheres hopping phenomena are not 
significant due to the absence of an energy scale, in molecular 
systems it is well known that the "MCT glass" is destabilized by 
activated processes, which slowly restore ergodicity. Indeed, in these 
systems the calorimetric glass transition temperature is located well 
below the MCT critical temperature.  In short-ranged attractive 
colloids, activated processes can be associated with thermal 
fluctuations of order $u_0$, which are able to break the bonds.  These 
processes generate a finite bond lifetime and destabilize the 
attractive glass.  Hence the effective $T_{cg}$ must lie below 
the value calculated on the basis of the ideal MCT attractive GL.
\bibitem{heavy} To give an estimation of this time in computer time, 
these simulations have required approximately 4 months in CPU time.
\end{thebibliography}
\end{document}